\documentclass[letterpaper, 10 pt, conference]{ieeeconf} 

\IEEEoverridecommandlockouts                              %

\title{\LARGE \bf
\thename: A Closed-Loop Simulation Framework For ICD Therapy
}
\author{Hannah Lydon$^{1}$, Milad Kazemi$^{1}$, Martin Bishop$^{2}$ and Nicola Paoletti$^{1}$%
\thanks{$^{1}$Hannah Lydon, Milad Kazemi and Nicola Paoletti are with the Department of Informatics, King's College London, London, UK. Corresponding author:
        {\tt\small hannah.lydon@kcl.ac.uk}}%
\thanks{$^{2}$Martin Bishop is with the Department of Biomedical Engineering, St. Thomas's Hospital, King's College London, London, UK}%
}

\usepackage{cite}
\usepackage{amsmath,amssymb,amsfonts}
\usepackage{graphicx}
\usepackage{subcaption}
\usepackage{textcomp}
\usepackage{xcolor}
\usepackage[hyphens]{url}
\usepackage{breakurl}
\usepackage{hyperref}
\usepackage{bm}
\usepackage[utf8]{inputenc}
\let\labelindent\relax
\usepackage{enumitem}
\usepackage{pifont} %
\usepackage{multirow}  %
\usepackage{comment}  %
\usepackage{algorithm}
\usepackage[indLines=false]{algpseudocodex}
\usepackage[normalem]{ulem}
\def\BibTeX{{\rm B\kern-.05em{\sc i\kern-.025em b}\kern-.08em
    T\kern-.1667em\lower.7ex\hbox{E}\kern-.125emX}}

\newcommand{\thename}{\textit{SimICD}}

\begin{document}

\maketitle
\thispagestyle{empty}
\pagestyle{empty}

\begin{abstract} 
Virtual studies of ICD behaviour are crucial for testing device functionality in a controlled environment prior to clinical application. Although previous works have shown the viability of using \textit{in silico} testing for diagnosis, there is a notable gap in available models that can simulate therapy progression decisions during arrhythmic episodes. This work introduces \thename, a simulation tool which combines virtual ICD logic algorithms with cardiac
electrophysiology simulations in a feedback loop, allowing the progression of ICD therapy protocols to be simulated for a range of tachy-arrhythmia episodes. Using a cohort of virtual patients, we demonstrate the ability of \thename\ to simulate realistic cardiac signals and ICD responses that align with the logic of real-world devices, facilitating the reprogramming of ICD parameters to adapt to specific episodes. 

\end{abstract}

\section{Introduction}
Implantable Cardioverter Defibrillators (ICDs) are essential tools for terminating potentially fatal rapid arrhythmias (tachy-arrhythmias) in high-risk patients using a range of pacing and electric shock therapy interventions. Whilst ICDs have proven effective in improving patient outcomes~\cite{ezekowitz_implantable_2003}, they are susceptible to misdiagnoses and ineffective therapy delivery~\cite{GOLD2012370, santini_prospective_2010}, highlighting the need for robust testing of ICD decision algorithms prior to clinical application.

Such vital testing can be conducted through \textit{in silico} experiments, which have been playing an increasingly significant role in clinical research by enabling safer, more comprehensive testing without invasive procedures whilst also allowing researchers to evaluate novel algorithms prior to use in real patient devices. Whilst previous work has shown the viability of using virtual devices for diagnoses of arrhythmia episodes~\cite{jiang_-silico_2016,jang2018computer}, these attempts only consider open-loop signals and thus cannot reproduce the effects of the device's therapy on the patient's cardiac electrophysiology (EP). Hence, a significant research gap remains in modelling the interactions between realistic virtual devices and advanced cardiac EP simulations in a closed-loop fashion, capable of both prescribing and simulating therapy interventions~\cite{trayanova_computational_2024, bhagirath_bits_2024}. 

 \begin{figure}
  \centering
  \includegraphics[width=\columnwidth]{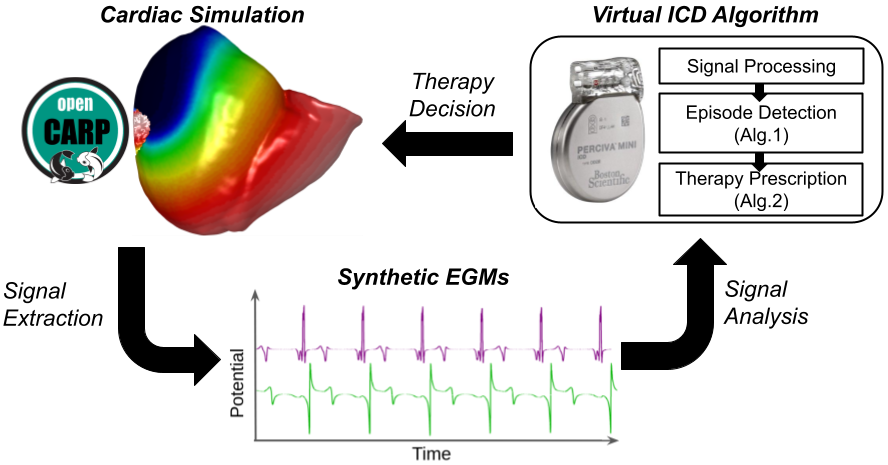}
  \caption{Visualization of \thename\ simulation pipeline. Cardiac arrhythmia episodes are modelled in OpenCARP \cite{plank_opencarp_2021}, the signals are extracted for analysis in the virtual ICD, and therapy decisions are relayed to the simulator.}
  \label{fig:Closed Loop}
\end{figure}

We address this gap by proposing \thename, an automated closed-loop simulation environment (represented in Fig. \ref{fig:Closed Loop}) that utilises the abilities of open-source cardiac modelling software and the algorithms of one of the leading device manufacturers to model therapy interventions during simulated cardiac episodes.

\paragraph*{Related work} We build on previous works~\cite{jiang_-silico_2016, paoletti_synthesizing_2019, krish_synthesizing_2023} for the encoding of the Boston Scientific ICD algorithm but significantly expand the algorithm to be in line with the latest available device manuals and support the ability for the virtual ICD to perform live signal monitoring, prescribe therapy and perform re-detection. Our work also leverages some of the methods for modelling idealized infarcts and re-entrant circuits~\cite{monaci_-silico_2020, monaci_automated_2021} but with a different research goal, namely studying closed-loop ICD therapy. The modelling of ATP therapy using the resources available in openCARP was primarily inspired by \cite{qian_-silico_2021} and the heart meshes, ICD electrode placement and the stimulus sites used in the patient simulations were leveraged from previous cardiac modelling investigations \cite{gemmell_computational_2020, plancke_generation_2019}.

In summary, our key contributions are as follows:
\begin{itemize}
    \item We present \thename, a novel closed-loop simulation environment that enables realistic \textit{in-silico} evaluation of multiple ICD therapies and several relevant cardiac episodes. Within this framework, we achieve:
    
        \item Implementation of a realistic virtual ICD model using the latest device manuals from Boston Scientific \cite{BostonScientific2020}, one of the leading device manufacturers, 
        \item Generation of a broad cohort of virtual patients for ICD evaluations using state-of-the-art cardiac EP modelling software \cite{plank_opencarp_2021}.
  
\end{itemize}

The codebase for the \thename\ tool is publicly available at: \href{https://github.com/janet-9/SimICD}{https://github.com/janet-9/SimICD}

\section{Background} \label{sec:background}

\subsection{ICD Functionality} \label{ssec: ICD function}
ICDs monitor the electric signals in the heart via leads embedded within the cardiac tissue and a pulse generator, also acting as a far-field sensor, placed in the torso. In a dual chamber device, these signals are traces from the atria, ventricles and the far-field signal and in a single chamber device, only the ventricular and far-field signals are measured. We focus on a single chamber device in this work as we are concerned with modelling arrhythmias of a ventricular origin.

The recorded signals are processed to form intracardiac electrograms (EGMs) that are analysed using internal device logic to diagnose potentially fatal arrhythmic episodes. These are typically classified as either Ventricular Tachycardia (VT), heart rates above 100 bpm, or Ventricular Fibrillation (VF), usually heart rates above 150 bpm and characterised by a lower amplitude, irregular rhythm~\cite{samie_mechanisms_2001}. VT, whilst considered the less severe of the two types, has the potential to develop into the (higher-risk) VF condition, and both episodes can result in sudden cardiac death if left untreated.

A common pacing therapy protocol used by ICDs to treat these arrhythmias is Anti-Tachycardia Pacing (ATP): rapid low-voltage pulses delivered from an electrode embedded within cardiac tissue designed to interrupt the mechanisms of the arrhythmia to restore a healthy heart rate (known as normal sinus rhythm or NSR, which is typically defined as a rhythm with a bpm between 60 - 100 \cite{spodick_operational_1992}).

If this intervention is unsuccessful, the device can proceed to deliver high-energy shocks, delivered via a current exchange between an electrode within the tissue and the pulse generator of the ICD, which increases the likelihood of termination but can cause discomfort to the patient and, if used excessively, have shown a correlation with worsened long-term patient outcomes \cite{li_significance_2016}. As such, the majority of devices are designed only to treat episodes that are potentially fatal and use the lower energy therapy options first before escalating the intensity.

The synthetic EGMs used for analysis by the virtual ICD in \thename\ were generated using the state-of-the-art source cardiac modelling software OpenCARP \cite{plank_opencarp_2021}, which allows for flexible implementation of a variety of cardiac meshes and electrophysiology models. 

\subsection{Electrophysiology Model} \label{ssec: EP Model}
For the cardiac EP simulations, we used various models of a human bi-ventricular (BiV) system to simulate both healthy and arrhythmic episodes. During all simulations, conduction in the myocardium was modelled using the monodomain model. This model, defined by equations \ref{eq:monodomain1}, \ref{eq:monodomain2}, describes propagation via a single parabolic reaction diffusion equation under the assumption that the intra- and extracellular spaces have proportional conductivities. It was chosen compared to the more complex bidomain model \cite{henriquez_simulating_1993}, which models the intra- and extracellular spaces separately because, due to its lower computational cost. The monodomain model, despite the reduced physiological accuracy, was found to have sufficient accuracy in simulating EGMs as it allowed the recovery of the extracellular potential, $\phi_{e}$, (the signal measured by ICD electrodes) by performing numerical integration at specific locations \cite{openCARP-paper}.

\begin{equation}
    \nabla \cdot \bm{\sigma_{m}} \nabla V_m = \beta (I_{m} + I_{stim}). 
    \label{eq:monodomain1}
\end{equation}

\begin{equation}
      \bm{\sigma_{m}} = \frac{\bm{\sigma_i}\bm{\sigma_e}}{\bm{\sigma_i} + \bm{\sigma_e}}. 
      \label{eq:monodomain2}
\end{equation}

In these equations, the tensor matrices \textit{\bm{$\sigma_{i}$}}, \textit{\bm{$\sigma_{e}$}} and values of \textit{$I_{i}$}, \textit{$I_{e}$} describe the conductivities and current densities of the intra and extracellular regions respectively, \textit{\bm{$\beta$}} represents the tissue level surface to volume ratio, $\phi_e$ represents the extracellular potential,  \textit{$I_{stim}$} is an external stimulus current, and \textit{$I_{m}$} is the transmembrane current, which itself can be described by equation (\ref{eq:transmembrane_current}), in which \textit{$C_{m}$} represents the cell membrane capacitance. 

\begin{equation}
  I_{m} = C_m \frac{\partial V_m}{\partial t} + I_{ion} - I_{stim}. 
    \label{eq:transmembrane_current}
\end{equation}

\section{ICD Modelling} \label{sec:methods}

The virtual ICD model includes sensing and decision logic algorithms based on the latest device manuals from Boston Scientific~\cite{BostonScientific2020}, implemented in Matlab. The key features of this virtual device are the availability of separated detection zones based on heart rate, the use of extracted features from EGMs to prescribe therapy, and the introduction of a re-detection algorithm that ensures that therapy interventions follow a prescribed progression protocol. 

\subsection{ICD Algorithms} \label{icd_algorithms}
The ICD logic comprises three main stages: 1) signal processing and feature extraction, 2) episode detection, and 3) therapy prescription. 
In particular, the first stage extracts the key features of the EGM signals, including the ventricular rate --- calculated using the ventricular period, i.e., the time between two consecutive measured beats --- and the vector timing and correlation (VTC) score, i.e., variation between the morphology of the measured signal and a (healthy) NSR template for that patient. 

In the second stage, these features are then analysed in moving windows of ten consecutive beats to detect if the patient's rhythm indicates a sustained episode in one of the four tachy-arrhythmia zones (VT1, VT, VF1, VF), characterised by the average cycle length recorded in the signal \cite{seifert_tachycardia_2013, noauthor_icd_nodate}. The splitting of the zones into slower (VT1, VF1) and faster regions is designed to allow less aggressive therapies to be prescribed for less severe (slower) episodes. More specifically, a round of ATP, known as Quick Convert (QC) ATP can be applied in the VF1 zone whereas episodes detected in the VF zone are treated with the delivery of a biphasic shock as the first therapy option \cite{noauthor_icd_nodate}. The detection logic is summarised in Algorithm~\ref{alg:epdetect} (executed in parallel once for each zone). For $\mathsf{zone \in \{VT1, VT, VF1, VF\}}$, the algorithm maintains a state $ (\mathsf{in_{zone}}, \mathsf{t_{zone}})$, where $\mathsf{in_{zone}}$ indicates whether it has entered a particular detection zone, which is triggered after 8+ beats out of 10 are faster than the detection threshold for that zone, $\mathsf{th_{zone}}$ (see lines 1-5); and $\mathsf{t_{zone}}$ is a clock keeping track of the time spent in that zone. When $\mathsf{t_{zone}}$ exceeds the device-set threshold $\mathsf{dur_{zone}}$, then the episode is classified as persistent and therapy is prescribed. 

Then, the particular therapy decision, see Algorithm~\ref{alg:prescription}, depends on the detection zone and how many (unsuccessful) ATP attempts have been made already (parameter $\mathsf{Tcount_{zone}}$) out of the available attempts (parameter $\mathsf{maxT_{zone}}$). We note that in the initial detection (no previous interventions have been made) for the two VT zones, an additional check is performed involving the VTC score to ensure that therapy is given only if the signal morphology does not correlate with the NSR template. This check is designed to inhibit interventions for tachycardias that do not originate in the ventricles \cite{noauthor_icd_nodate} and is not performed when prescribing therapy in re-detection mode as per the prescription from the device manual. The algorithm also returns two parameters used to configure the ATP therapy, i.e., the average (using the last four measured beats) and the final measured ventricular period in the detected episode, denoted $\mathsf{avgVperiod}$ and $\mathsf{Vtime}$ respectively. Finally, therapy is initiated in the highest-risk zone, which is the first to meet the criteria for intervention.

\begin{algorithm}
\caption{\textbf{Determine a sustained episode in any zone} \\ 
\textbf{Inputs:} 
Latest $10$ ventricular periods ($\mathsf{VPeriods}$);  
Current detection state ($\mathsf{in_{zone}}$, $\mathsf{t_{zone}}$);  
ICD parameters: Detection and duration thresholds ($\mathsf{th_{zone}}$,  $\mathsf{dur_{zone}}$) \\
\textbf{Output:} Updates next state $\mathsf{(in^{'}_{zone}, t^{'}_{zone})}$ and returns $\mathsf{True}$ iff a sustained episode is detected.}
\label{alg:epdetect}

\begin{algorithmic}[1]
    \LComment{Count how many individual periods, $V_{period}$, are shorter than $\mathsf{th_{zone}}$}
    \State $\mathsf{fastCount\gets \sum_{V_{period} \in VPeriods} \mathbf{1}(V_{period}<th_{zone})}$ 

    \LComment{Enter zone after $8+$ fast periods, including the last}
    \If{$\mathsf{fastCount \geq 8}$ \textbf{and} $\mathsf{VPeriods[\textbf{end}]<th_{zone}}$} 
        \State $\mathsf{in^{'}_{zone} \gets True}$
    \EndIf

    \LComment{Update clock if episode persists ($6+$ fast periods)}
    \If{$\mathsf{in_{zone}}$}        
        \If{$\mathsf{fastCount \geq 6}$ \textbf{and} $\mathsf{Vbeats[\textbf{end}]<th_{zone}}$} 
            \State $\mathsf{t^{'}_{zone} \gets t_{zone} + Vbeats[\textbf{end}]}$
        \Else
            \State $\mathsf{(in^{'}_{zone},t^{'}_{zone}) \gets (False,0)}$
        \EndIf
    \EndIf 

    \LComment{Episode is sustained if clock exceeds threshold}
    \State \Return ($\mathsf{t_{zone}\geq dur_{zone}}$)
\end{algorithmic}
\end{algorithm}

\begin{algorithm}
\caption{\textbf{Therapy prescription for sustained episodes} \\  
\textbf{Inputs:}  
$\mathsf{VPeriods}$ from the sustained episode flagged in Algorithm \ref{alg:epdetect};  
Absolute time of last ventricular period ($\mathsf{Vtime}$);  
Detection zone of the sustained episode ($\mathsf{zone}$);  
Latest 10 correlation scores ($\mathsf{VTCs}$);  
Correlation threshold ($\mathsf{VTC_{th}}$);  
Flag for an initial episode ($\mathsf{initial}$);  
Number of previous therapy interventions ($\mathsf{Tcount_{zone}}$), and maximum interventions allowed per zone ($\mathsf{maxT_{zone}}$)  \\
\textbf{Output:}  
Updates number of interventions ($\mathsf{Tcount^{'}_{zone}}$) and returns therapy decision ($\mathsf{therapy}$) with corresponding parameters ($\mathsf{therapyParams}$)}

\label{alg:prescription}

\begin{algorithmic}[1]
    \LComment{Average ventricular period of the detected episode}
    \State $\mathsf{avgVperiod} \gets \frac{1}{4}\sum_{i=7}^{10} \mathsf{VPeriods}[i]$

    \If{$\mathsf{zone = VF1}$ \textbf{and} $\mathsf{Tcount_{VF1} < max_{VF1}}$}
        \LComment{Trigger QC ATP if attempts are available}
        \State  $\mathsf{therapy} \gets \mathsf{QCATP}$ 
        \State  $\mathsf{therapyParams} \gets \mathsf{(avgCL, Vtime)}$
        \State $\mathsf{Tcount^{'}_{zone}}\gets \mathsf{Tcount_{zone}}+1$

    \ElsIf{ ($\mathsf{zone = VT}$ \textbf{or} $\mathsf{zone = VT1}$) \textbf{and} $\mathsf{Tcount_{zone} < maxT_{zone}}$}
        \LComment{Trigger ATP therapy; for initial episodes, check if rhythm is uncorrelated}
        \State $\mathsf{corrCount\gets \sum_{score \in VTCs} \mathbf{1}(score > VTC_{th})}$
        \If{$\mathsf{\neg initial}$ \textbf{or} $\mathsf{corrCount \leq 3}$}
            \State  $\mathsf{therapy} \gets \mathsf{ATP}$
            \State  $\mathsf{therapyParams} \gets \mathsf{(avgCL, Vtime)}$
            \State $\mathsf{Tcount^{'}_{zone}}\gets \mathsf{Tcount_{zone}}+1$
        \Else
            \LComment{Inhibit therapy for correlated initial episodes}
            \State  $\mathsf{therapy} \gets \mathsf{Inhibit}$; 
            $\mathsf{therapyParams} \gets \mathsf{nil}$
        \EndIf

    \Else 
        \LComment{If (QC) ATP attempts exhausted or in VF zone}
        \State  $\mathsf{therapy} \gets \mathsf{Shock}$; 
        $\mathsf{therapyParams} \gets \mathsf{nil}$
    \EndIf

    \State \Return $(\mathsf{therapy,therapyParams})$

\end{algorithmic}
\end{algorithm}

\subsection{Therapy Simulations} \label{ssec:atp}
Based on the ICD technical specifications, we implemented two ATP schemes: burst (equally spaced pulses) and ramp (pulses that gradually decrease in cycle length). As per the default settings laid out the Boston Scientific manual \cite{noauthor_icd_nodate} the default therapy progression pathway was set to use the burst scheme for the first round of ATP and the ramp scheme for the second round when treating VT and VT1 episodes. The virtual electrodes were designed to replicate the electrodes in the Boston Scientific single chamber ICD \cite{BostonScientific2020}. These include a left pectoral can and a right ventricular lead with a sensing coil, a ring, and a tip electrode. The sensing electrodes were modelled as single points embedded within the tissue as this was found to be sufficient to record the local potential, whilst the pacing electrode, the tip of the right ventricular lead, had an approximate volume of $5\times5\times5$ mm$^{2}$ to more accurately capture the spatial influence of the therapy stimulus.

Our simulation framework enables personalisation of the scheme to a fine degree (and in line with device manuals) by allowing the following therapy parameters to be adjustable: the number of pulses, the coupling time (the time between the last measured beat in the episode and the start of the delivery), the pacing intervals, and the interval decrement during the ramp scheme. The default coupling time and pacing interval were set to the nominal value of 81$\%$ of the average cycle length of the episode (calculated from the last four measured intervals). As per the device manual, the parameters for QC ATP, only available in the VF1 zone, were fixed at 88$\%$ of the average cycle length and could only be delivered once in an episode. The full list of programmable therapy parameters can be seen in Table \ref{tab:ATP_param_table}.

\subsection{Closed Loop Automation} \label{automation}

In \thename, the virtual ICD model and the cardiac EP model run in parallel. The ICD starts in idle mode, waiting for EGM signals from the EP model. The EP model transforms extracellular potential signals read by its virtual electrodes into an EGM signal. The signals are exported in output files which are continuously monitored and processed by the ICD model. Only after receiving an EGM for a full detection window, the ICD model can trigger its episode detection and therapy prescription pipeline. If therapy is required, \thename\ triggers the execution of the EP model with the relevant therapy. 

In particular, our approach saves multiple checkpoints for the EP model state so that when switching from a no-therapy to a therapy simulation (or vice versa), the simulation is restarted at the correct timestamp when therapy is prescribed/completed. This way, we do not need to synchronise the ICD and EP models explicitly, and they can execute asynchronously. \thename\ continues the detection and therapy delivery procedure until either the detection/re-detection algorithm does not trigger a therapy intervention or all therapy options have been exhausted for that episode.

\section{Virtual Patient Cohort} \label{virtual patients}

The virtual patient cohort was created using three distinct meshes and multiple pacing protocols to simulate a range of episode types, which included NSR and two varieties of ventricular tachycardia (VT): \textit{focal VT}, due to ectopic excitations and \textit{re-entrant VT},  due to scar related circuits forming in the tissue. The patient cohort is detailed in Table~\ref{tab:Patient_table}. As the only episodes considered in this work were those of ventricular origin, only the ventricles were included in the mesh model, as has been done in previous modelling studies of ventricular arrhythmias~\cite{plancke_generation_2019, qian_-silico_2021, monaci_non-invasive_2023}.

 \begin{table}[h!]
    \centering
    \begin{tabular}{|c|c|c|c|c|}
        \hline
        \multirow{2}{*}{\textbf{Patient No.}} & \multirow{2}{*}{\textbf{Mesh}} & \multicolumn{3}{|c|}{\textbf{Available Episode Type}} \\ \cline{3-5} 
         & & NSR & Focal VT & Reentrant VT \\ \hline
        0     &  BiV (no scarring)   &  \ding{51} & \ding{55} & \ding{55}             \\ \hline
        1     &  BiV (no scarring)   &  \ding{51} & \ding{51} & \ding{55}             \\ \hline
        2     &  BiV (LV scarring)   &  \ding{51} & \ding{51} & \ding{51}             \\ \hline
        3     &  BiV (RV scarring)  &  \ding{51} & \ding{51} & \ding{51}             \\ \hline
    \end{tabular}
    \caption{Virtual patient cohort summary. \textit{LV} and \textit{RV} denote the left and right ventricle respectively. Patient 0 represents a structurally normal healthy patient baseline.}
    \label{tab:Patient_table}
\end{table}

\begin{figure}
    \centering
        \includegraphics[width=0.6\linewidth]{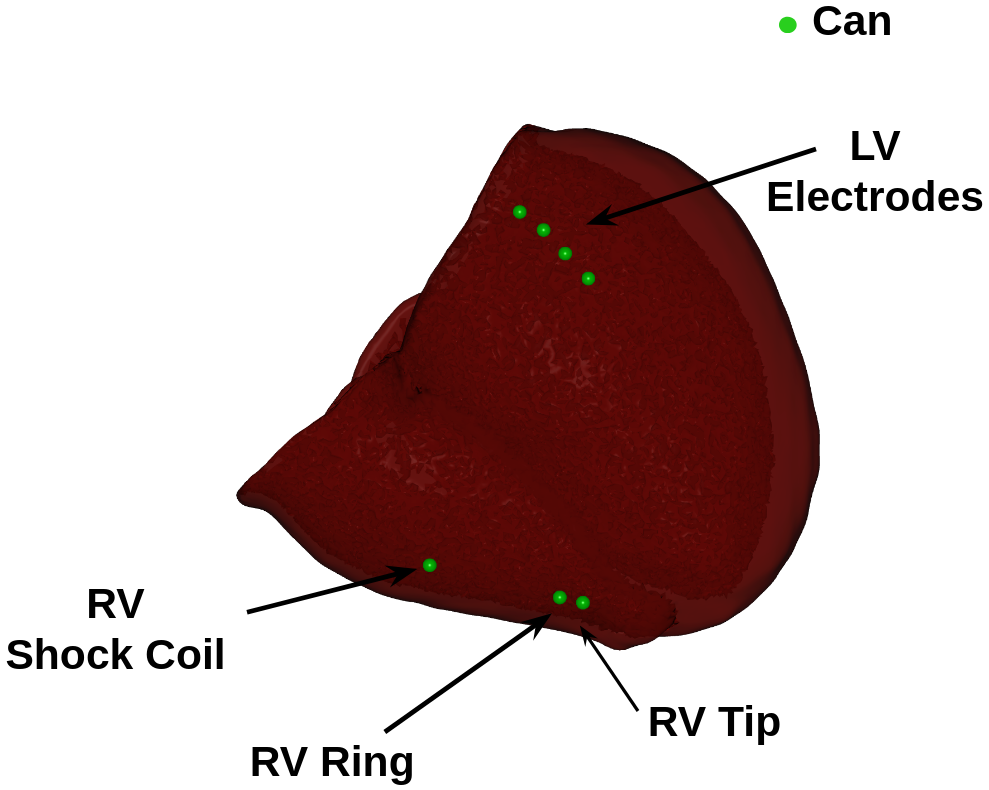}
    \caption{Visualisation of the BiV mesh and the ICD electrodes}
    \label{fig:electrodes}
\end{figure}

All patient meshes were derived from a single BiV model, which was generated using computed tomography (CT) data from real patients and has previously been used in studies of simulated ventricular arrhythmia episodes~\cite{plancke_generation_2019, gemmell_computational_2020}. A representation of the structurally normal BiV mesh and the virtual ICD electrodes can be seen in Fig.~\ref{fig:electrodes}.

\subsection{Mesh Preparation}\label{meshprep} The tenTusscher-Panfilov cell model \cite{ten_tusscher_model_2004} was used in all of the cardiac simulations. In the initialisation phase, a single cell was paced at 1.25Hz (75bpm) for 100 cycles to achieve a stable limit cycle and the conductivities required to achieve ventricular myocyte conduction velocities based on literature data \cite{caldwell_three_2009} were tuned using a strip of tissue with the same resolution as the full mesh. The stable cell states and calculated conductivity values were inserted into the BiV meshes, which were then paced from the apex, and the conductivities were further tuned to achieve stability at the tissue level. 

\subsection{NSR Simulation}\label{nsr_gen} Models were paced from approximate sinus rhythm locations, as defined by \cite{gemmell_computational_2020}, with a strength of 450$\mu$Acm$^{-3}$ for 4 ms to achieve a basic cycle length of 800ms (75bpm), which was then used an initial state for future simulations.
During all episodes of simulated tachycardia in \thename\ , the sinus points of the mesh continue to be paced at the NSR rate of 75bpm so that successful termination of a tachycardia episode results in the restoration of a healthy rhythm as it would in a real-world ICD. 
 
\subsection{Focal VT simulation} \label{focalvts}
Episodes of focal VT were modelled using small stimulus sites to represent spontaneous ectopic beats. The locations of these sites were chosen to represent some common cases of focal VT in patients, and were modelled as a transmembrane stimulus with the same strength and duration as the sinus rhythm excitation sites. During simulations of focal VT episodes within the \thename\ environment, the number of ectopic beats in an episode, the cycle length, and the number of ectopic excitation episodes are all variable parameters that can be used to change the characteristics of the episode. 

\subsection{Re-entrant VT Simulation} \label{anatomical vts}
\begin{figure}
    \centering
    \begin{subfigure}{\columnwidth}
        \centering
        \includegraphics[width=\linewidth]{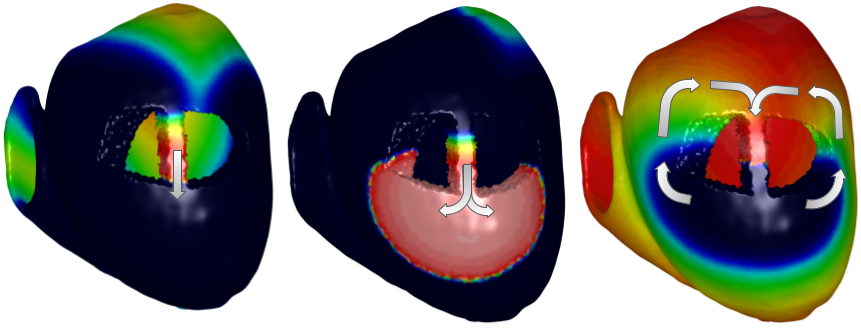}
        \caption{}
        \label{fig:LV_1}
    \end{subfigure}
    \hfill
        \begin{subfigure}{\columnwidth}
        \centering
        \includegraphics[width=\linewidth]{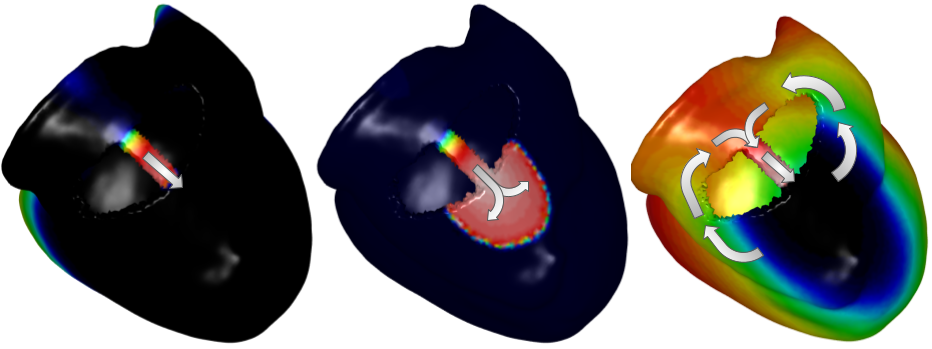}
        \caption{}
        \label{fig:LV_2}
    \end{subfigure}
    \hfill
        \begin{subfigure}{0.7\columnwidth}
        \centering
        \includegraphics[width=\linewidth]{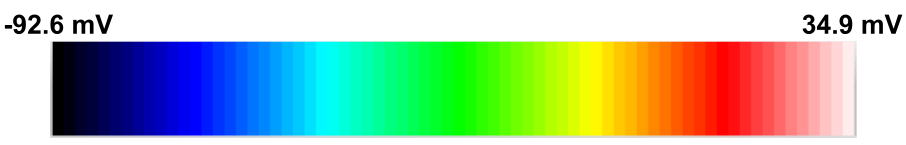}
        \label{fig:colourbar}
    \end{subfigure}
    \hfill
    \caption{Voltage maps of the stages of a re-entrant circuit during VT simulations in openCARP. \textit{a)} patient 2 and \textit{b)} patient 3. 
    \textit{From left to right:} wavefront travels down through the isthmus and recirculates around the non-conducting scar tissue to re-enter.}
    \label{fig:Reentrant_Stills}
\end{figure}

The episodes of re-entrant VT are simulated using idealised scars designed to facilitate a double loop re-entrant circuit, which is a common mechanism for re-entry \cite{kleber_basic_2004}. These scars are formed of an elliptical region of unexcitable tissue with a rectangular strip through the centre that acts as a slow conducting isthmus, allowing the excitation wavefront to propagate around the scar in a figure of eight patterns, as shown in Figure~\ref{fig:Reentrant_Stills}. The non-conducting regions are represented by missing volumes to both increase computational efficiency and enforce a sharp conduction barrier. Whilst it is possible that these missing volumes could affect the current density of the ICD therapy, the strength and duration of the stimulus pulses were found to be sufficient to excite the tissue and propagate across the mesh as expected.

Initial re-entrance was induced by reducing the conductivity at one end of the isthmus to create a unidirectional block and stimulating the surrounding tissue. Once the depolarization wavefront had started to enter the isthmus, the blocked region had its conductivity restored, and the excitation was then allowed to propagate around the scar - resulting in a self-sustained double loop re-entrant circuit. The cycle length of the re-entrant circuit can be altered when simulating VT episodes in \thename\ by scaling the conductivities within the isthmus relative the healthy tissue, and the loops can propagate in both antegrade and retrograde directions.

\section{Results} \label{sec:results}

\subsection{Non-sustained VT} \label{non-sustained episode}
\begin{figure}
  \centering
  \begin{subfigure}[b]{\columnwidth}
    \centering
    \includegraphics[width=0.9\columnwidth]{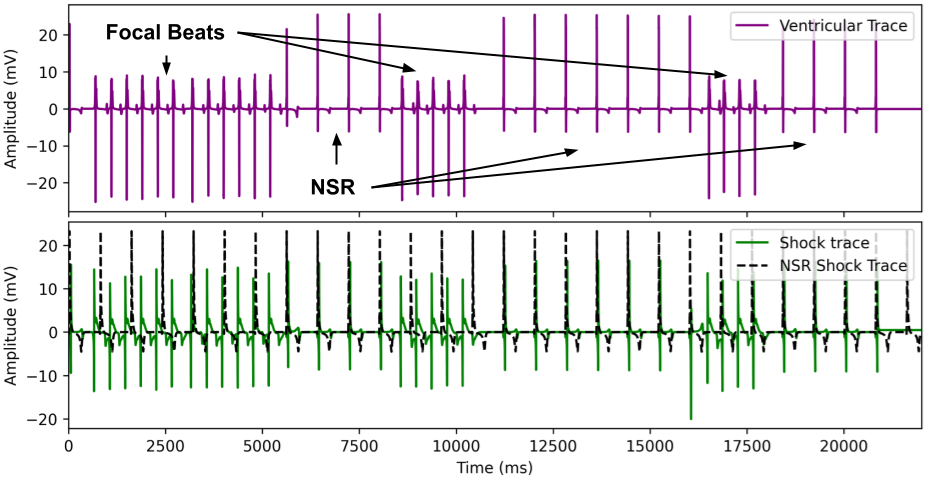}
    \caption{}
    \label{fig:RVOT_EGM}
  \end{subfigure}
  \begin{subfigure}[b]{\columnwidth}
    \centering
    \includegraphics[width=0.9\columnwidth]{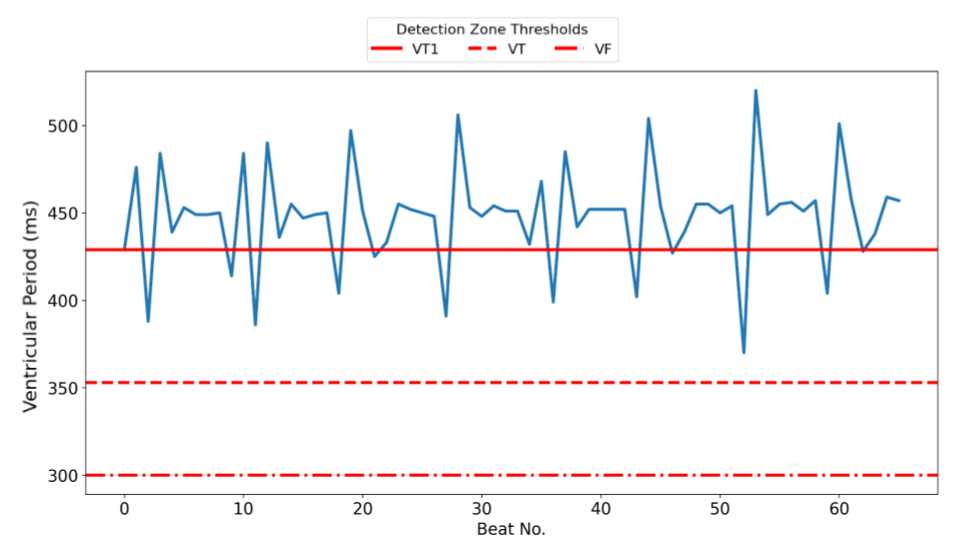}
    \caption{}
    \label{fig:Beat_ev_RVOT}
  \end{subfigure}
  \caption{Results of the RVOT episode simulation in patient 1. \textit{a)} The EGM of the episode, with the NSR template shown for comparison for the shock trace. \textit{b)} The evolution of the ventricular periods (blue), with the detection zones shown.}
  \label{fig:RVOT_analysis}
\end{figure}

The focal VT episode that was used as an example simulation was the common and typically benign non-sustained arrhythmia called RVOT (Right Ventricular Outflow Tract) VT \cite{lin_right_2014}, in which ectopic beats occur in the top of the right ventricle. Episodes of this nature are typically not sustained long enough to warrant intervention from an ICD. To test the response of the \thename\ to this type of signal, a simulation of RVOT in patient 1 (containing focal bursts ranging between 120 - 150 bpm for 8 to 10 pulses) was monitored over the course of a 30-second episode simulation. The EGMs recorded by \thename\ during the monitoring process and the evolution of the recorded ventricular periods are shown in Fig.~\ref{fig:RVOT_analysis}. 
 
The measured ventricular periods in this episode were very unstable, and even though parts of the episode were classified as being within the VT1 detection zone, the algorithm did not remain in that zone long enough to meet the duration threshold. Therefore, therapy intervention was inhibited. This behaviour mirrors the expected behaviour of the ICD in this type of non-sustained arrhythmia and thus shows the viability of \thename\ to reproduce realistic responses to short-lived arrhythmic episodes. 

\subsection{Sustained VT episodes and Parameter Adjustment}\label{ssec:susepisodes}

 \begin{figure}
  \centering
  \begin{subfigure}{\columnwidth}
    \centering
    \includegraphics[width=0.7\columnwidth]{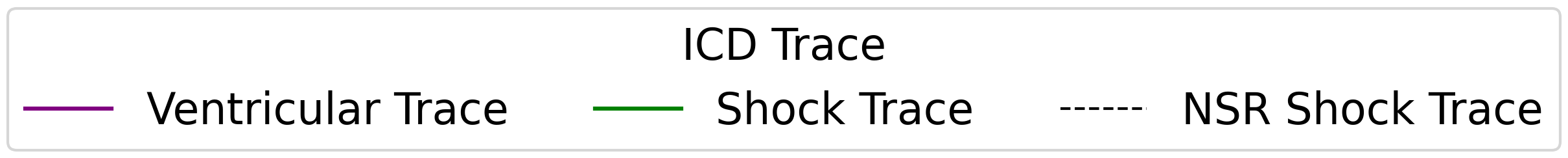}
  \end{subfigure}
  \begin{subfigure}{\columnwidth} 
    \centering
    \includegraphics[width=\columnwidth, height=4cm]{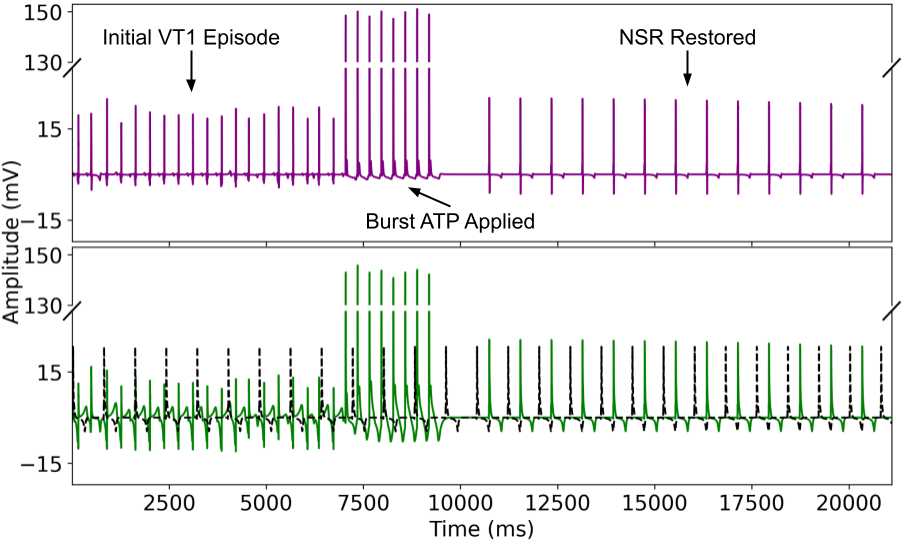}
    \caption{}
  \end{subfigure}
  \begin{subfigure}{\columnwidth} 
    \centering
    \includegraphics[width=\columnwidth, height=4cm]{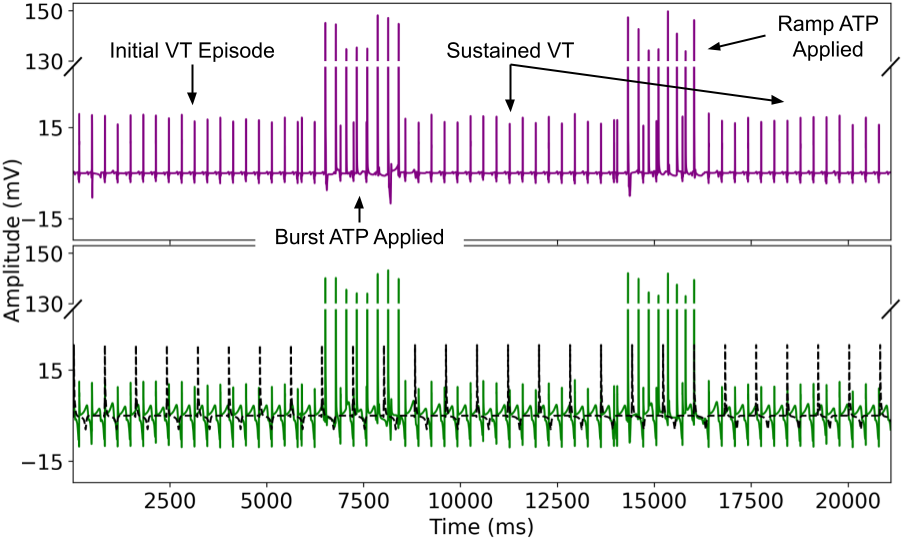}
    \caption{}
  \end{subfigure}
  \caption{EGM traces for the \textit{(a)} VT1 and \textit{(b)} VT  episodes and therapy interventions for Patient 2 (left ventricle scarring).}
  \label{fig:LV_analysis}
\end{figure}

This example demonstrates the ability of \thename\ to follow therapy protocols in response to sustained tachy-arrhythmia episodes and consists of two parts. The delivery of ATP therapy using the nominal programming parameters was assessed against a slow and a fast VT episode, and then the ATP parameters were adjusted to prescribe therapy to a previously unsuccessfully treated episode to improve the outcome. 

\subsubsection{Nominal Parameters}
The conductivities within the isthmus of the meshes for patients 2 and 3 were adjusted to simulate two distinct re-entrant VT episodes. For patient 2, the isthmus conductivities were scaled to 20$\%$ of the myocardium for the VT1 episode and 25$\%$ for the VT episode and for patient 3 the conductivities were scaled to 15$\%$ and 18$\%$ for the VT1 and VT episodes respectively. When using the default ATP parameters and protocol described in Section \ref{ssec:atp}, we found that for both patients the VT1 episode was terminated after a single round of burst ATP, but the episodes of VT persisted despite delivering both available rounds of ATP therapy. An example of the EGMs showing the VT and VT1 episodes for patient 2 can be seen in Fig.~\ref{fig:LV_analysis}. 

\begin{table}[h!]
    \centering
    \large
    \resizebox{\columnwidth}{!}{
        \begin{tabular}{|c|c|c|c|c|c|}
            \hline
            \multirow{2}{*}{\textbf{\textbf{ATP Parameter}}} & \multirow{2}{*}{\textbf{\textbf{VF1 ($<$300 ms)}}} & \multicolumn{3}{c|}{\textbf{VT ($<$ 353 ms)}} & \multirow{2}{*}{\textbf{VT1 ($<$ 429 ms)}} \\
            \cline{3-5}
            & & Default & New (P2) & New (P3) &  \\ \hline
            Pulse Interval (\%) & 88 & 81 & 88 & 88 & 81\\ \hline
            Coupling Interval (\%) & 88 & 81 & 88 & 88 & 81\\ \hline
            No. Pulses & 8 & 8 & 12 & 8 & 8 \\ \hline
            \parbox[t]{3cm}{\centering Ramp\\ Decrement (ms)} & N/A & 10 & 5 & 5 & 10 \\ \hline
        \end{tabular}}
    \caption{Parameters for ATP therapy in each therapy zone. For the VT zone, the default and the adjusted values following testing for patients 2 and 3 (P2, P3) are shown.}
    \label{tab:ATP_param_table}
\end{table}

These results demonstrate two important facts: 1) our virtual ICD correctly responds to the recorded signals from the cardiac simulations and 2) there are limitations in using the nominal ATP parameters and protocols to treat all tachycardia episodes, an idea that has been suggested and discussed in various clinical papers~\cite{de_maria_antitachycardia_2017, qian_-silico_2021, awad_arrhythmogenicity_2023, gulizia_randomized_2009}. In particular, it has been shown that traditional ATP therapy shows reduced effectiveness in treating faster VT episodes~\cite{santini_prospective_2010}, a finding that we replicate in our experiments using \thename, as described next.

\subsubsection{Parameter adjustments}

\begin{figure}
  \centering
    \includegraphics[width=\columnwidth]{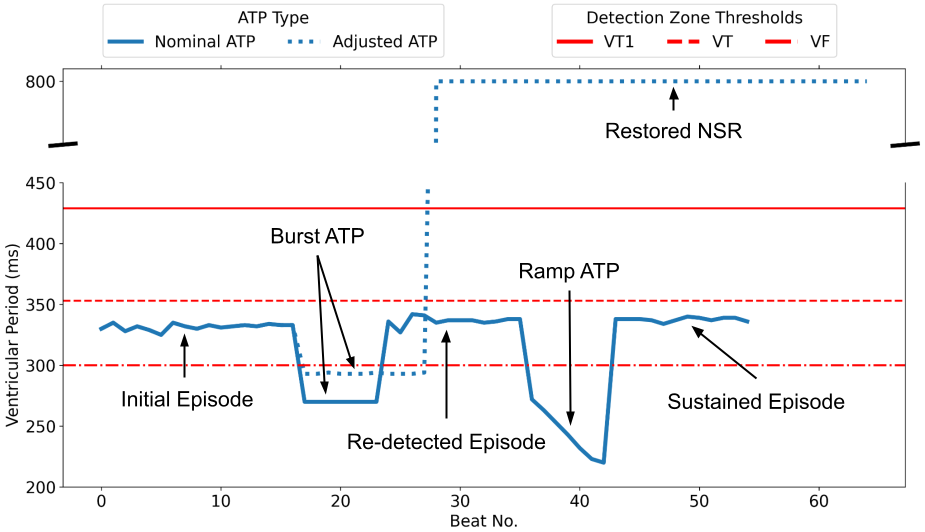}
  \caption{The evolution of the ventricular periods during simulations using adjusted vs. nominal ATP parameters for the treatment the VT episode in Patient 2. The traces end when either the episode is deemed to be terminated or therapy options have been exhausted.}
  \label{fig:LV_analysis_3}
\end{figure}

\begin{figure}
  \centering
  \begin{subfigure}{\columnwidth}
    \centering
    \includegraphics[width=0.7\columnwidth]{Figures/Legend.png}
  \end{subfigure}
  \begin{subfigure}{\columnwidth} 
    \centering
    \includegraphics[width=\columnwidth, height=4cm]{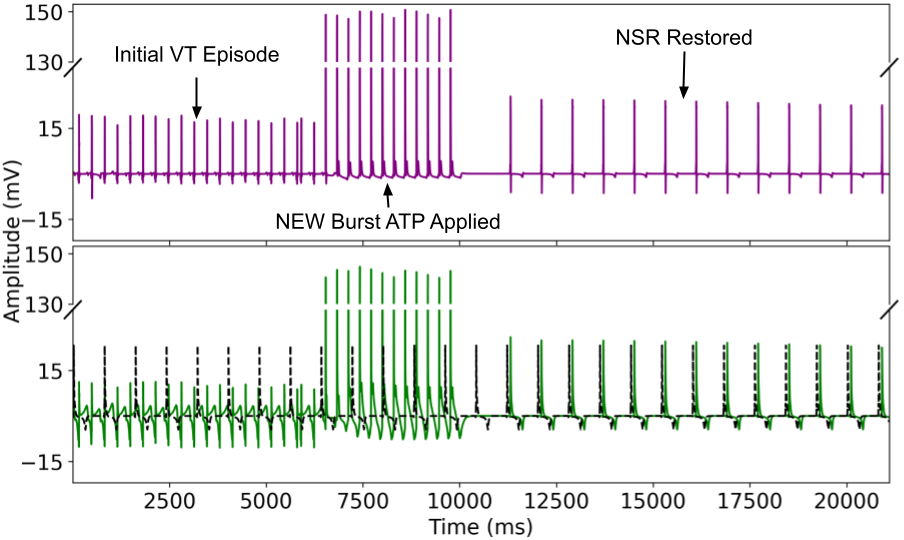}
    \caption{}
  \end{subfigure}
  \begin{subfigure}{\columnwidth} 
    \centering
    \includegraphics[width=\columnwidth, height=4cm]{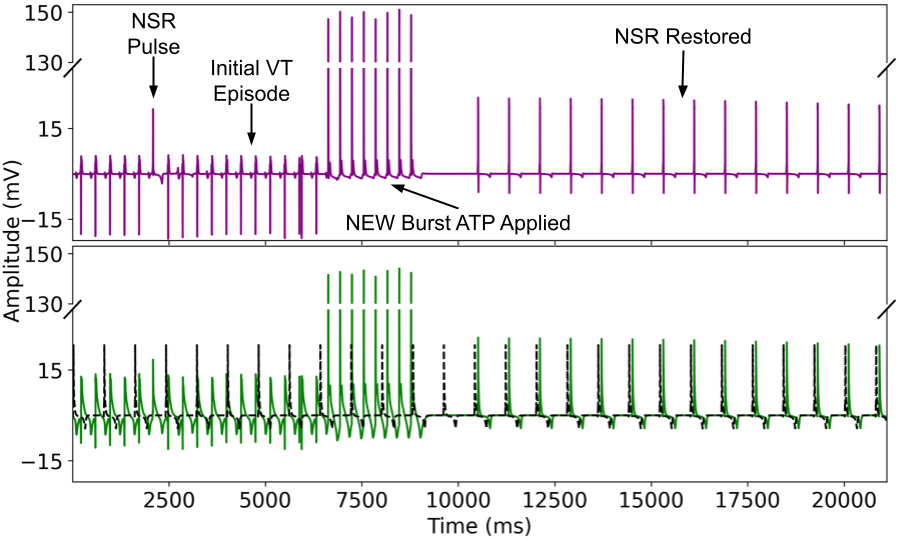}
    \caption{}
  \end{subfigure}
  \caption{EGM traces for the VT episodes in \textit{(a)} Patient 2 and \textit{(b)} Patient 3 using new ATP parameters, adapted to the specific episode.}
  \label{fig:New_ATP_EGMs}
\end{figure}

Our \thename\ framework allows for the ATP parameters to be adjusted in order to assess the therapy outcomes under different conditions. As an example, the VT episodes that persisted during the default ATP treatment, were tested using a range of new parameters, following the recommendations of from~\cite{de_maria_antitachycardia_2017}. We first increased the coupling and pulse intervals to 88$\%$ and then increased the number of burst pulses (within the recommended range of 8-15) until the episode was terminated. Following these tests, it was found that the increased pulse interval was sufficient to terminate the VT in patient 3 within 8 pulses, whilst 12 pulses were required to terminate the VT in patient 2. The adjusted parameters for the VT zone are shown in Table \ref{tab:ATP_param_table} and the results of using the new therapy parameters to treat the episode can be seen in Fig.~\ref{fig:LV_analysis_3} and Fig.~\ref{fig:New_ATP_EGMs} : demonstrating that by adjusting the parameters of the ATP, the VT episodes can be terminated after a single round of therapy.

\section{Conclusion}\label{sec:discussion}
This work introduces \thename, a novel closed-loop environment for modelling realistic interactions between a virtual human heart model and an accurate ICD algorithm that can detect tachy-arrhythmias, prescribe therapy based on the features of the episode and follow prescribed therapy protocols during sustained episodes. By performing simulation studies on a cohort of virtual patients, we demonstrated that \thename\ can accurately detect and prescribe therapy in accordance with device manuals and replicate the findings of various clinical and \textit{in silico} studies. \thename\ also has a high degree of flexibility for altering the ICD parameters to improve therapy outcomes. This shows its viability as a useful test-bed for studying ATP protocols and ICD programming strategies.

\section*{Acknowledgments}
 This work was funded by the UKRI EPSRC grant MCPS-VeriSec (EP/W014785/2) and British Heart Foundation (BHF) Project Grant PG/22/10871.

 \bibliographystyle{unsrt}

\bibliography{references}

\end{document}